\documentclass[a4paper,11pt]{article}
\pdfoutput=1 

\usepackage{jinstpub} 

\usepackage[T1]{fontenc} 
\usepackage{multirow}

\usepackage{graphicx}
\usepackage{color}
\usepackage{mathtools}
\usepackage[left]{lineno}
\usepackage{hyperref}
\usepackage{subfigure}
\usepackage{gensymb}
\usepackage{wrapfig}

\usepackage{natbib}

\title{R2D2 TPC: first Xenon results}

\author[a]{R.~Bouet}
\author[b]{J.~Busto}
\author[a,f]{V.~Cecchini}
\author[a]{C.~Cerna}
\author[a]{P.~Charpentier}
\author[c]{M.~Chapellier}
\author[d]{A.~Dastgheibi-Fard}
\author[a]{F.~Druillole}
\author[a]{C.~Jollet}
\author[a]{P.~Hellmuth}
\author[e]{M.~Gros}
\author[f]{P.~Lautridou}
\author[a,1]{A.~Meregaglia\note{Corresponding author}}
\author[e] {X.~F.~Navick}
\author[a]{F.~Piquemal}
\author[e] {F.~Popieul}
\author[a]{M.~Roche}
\author[g]{I.~Savvidis}
\author[a]{B.~Thomas}
\affiliation[a]{LP2I Bordeaux, Universit\'{e} de Bordeaux, CNRS/IN2P3, F-33175 Gradignan, France}
\affiliation[b]{CPPM, Universit\'{e} d'Aix-Marseille, CNRS/IN2P3, F-13288 Marseille, France}
\affiliation[c]{IJCLab, CNRS/IN2P3, Paris, France}
\affiliation[d]{LPSC-LSM, CNRS/IN2P3, Universit\'{e} Grenoble-Alpes, Modane, France}
\affiliation[e]{IRFU, CEA, Universit\'{e} Paris-Saclay, F-91191 Gif-sur-Yvette, France}
\affiliation[f]{SUBATECH, IMT-Atlantique, Universit\'{e} de Nantes, CNRS-IN2P3, France}
\affiliation[g]{Aristotle University of Thessaloniki, Thessaloniki, 54124 Greece}

\abstract{
Radial time projection chambers (TPC), already employed in the search for rare phenomena such as light Dark Matter candidate, could provide a new detection approach for the search of neutrinoless double beta decay ($\beta\beta0\nu$). The assessment of the performances of such a detector for $\beta\beta0\nu$ search is indeed the goal of the Rare Decays with Radial Detector (R2D2) R\&D. Promising results operating a spherical TPC with argon up to 1~bar have been published in 2021. Supplementary measurements were recently taken extending the gas pressure range up to 3~bar. In addition, a comparison between two detector geometries, namely spherical (SPC for spherical proportional counter) and cylindrical (CPC for cylindrical proportional counter), was performed. Using a relatively simple gas purification system the CPC detector was also operated with xenon at 1~bar: an energy resolution of 1.4\% full-width at half-maximum was achieved for drift distances up to 17~cm. Much lower resolution was observed with the SPC.  These results are presented in this article.}

\begin{document}
\maketitle
\flushbottom

\section{Introduction}

Initiated in 2018, the R2D2 R\&D program aims to develop a time projection chamber (TPC), based on a spherical  proportional counter (SPC), for the search for neutrinoless double beta decay ($\beta\beta0\nu$).\\
Such a detector was initially developed in CEA Saclay~\cite{Giomataris:2008ap}, primarily aiming to study low energy neutrino physics such as: neutrino oscillations, neutrino coherent elastic scattering and supernova neutrino detection~\cite{Giomataris:2003bp,Giomataris:2005fx}. The idea was to combine the possibility for good energy resolution, low energy threshold, low background capability and large target masses within a novel gaseous detector design. 
Within the R2D2 project, several aspects were studied in order to optimize the SPC for an unambiguous detection of the $\beta\beta0\nu$ process signal: two electrons in the MeV region.
 After initial results, and considering the proximity of the two concepts, a cylindrical proportional counter (CPC) was also recently studied. 

The SPC consists of a large grounded sphere with a central small spherical anode of $\sim$1~mm radius, which constitutes the only readout channel.
This simple design optimizes the volume to surface ratio, while allowing a low capacitance and high gain. Due to these  features, SPCs are already used by the NEWS-G collaboration~\cite{Gerbier:2014jwa,Arnaud:2017bjh,Savvidis:2016wei} for the search of light dark matter candidates in the mass range between 0.1 and 10~GeV, where the detection of single ionization electrons is routinely performed.
 The possibility of percent level energy resolution with a minimal material budget and large mass makes SPCs an appealing option in the search of other rare phenomena such as $\beta\beta0\nu$ decay.
This has previously been investigated in a $^{136}$Xe-filled detector at a pressure of 40 bar~\cite{Meregaglia:2017nhx} which suggested a sensitivity to the inverted mass hierarchy region, and influenced the R2D2 R\&D.\\
The initial goal was to establish an energy resolution at the level of 1\% full-width at half-maximum (FWHM) at 2.458~MeV, corresponding to the transition energy ($Q_{\beta\beta}$) of the $^{136}$Xe double beta decay.
The first results with a prototype built at LP2I Bordeaux and operated with a gas mixture of 98\% argon and 2\% methane (ArP2) validated the detector performance up to 1~bar ~\cite{Bouet:2020lbp}. Recent measurements, obtained with a new prototype of SPC and confirming the previous results up to a pressure of 3~bar, will be  presented in this paper. Furthermore, the first results obtained operating the detector with xenon up to 1~bar will also be discussed.

Several pitfalls could be encountered when operating SPCs in proportional mode at 40~bar: a high-voltage of a few tens of kV should be necessary, perfect electrical insulation between the central ball and the rod becomes essential, and the electronic noise (due to the decoupling capacitance) could increase and deteriorate the energy resolution. Furthermore, related to the peculiarity of the xenon gas, the very weak electric field near the cathode ($1/R^2$ dependence) could alter the collection of the charges in the presence of electronegative impurities, even with low contamination at the parts-per-billion (ppb) level.\\
The adoption of a cylindrical geometry would make it possible to circumvent these difficulties, in particular using a negatively polarized cathode (providing better voltage resistance) and a central grounded wire (resulting in a noise independent of the HV). The higher electric field at large radii (dependence as $1/R$) is also expected to relax the constraints on gas purity.\\
A CPC prototype was built at SUBATECH Nantes and operated at LP2I Bordeaux both in ArP2 and in xenon up to 1 bar showing that a resolution at the percent level can be achieved. Such results will be discussed in the paper.

\section{Detectors}

The working principle of a SPC or a CPC relies on the fact that particles traversing the gas ionize it, with the ionization electrons drifted toward a central anode (a sphere or wire in a SPC or CPC respectively) by an electric field. Once the electrons get close to the anode, depending on the high-voltage applied, they enter the avalanche region where they are multiplied and collected on the anode. The multiplication of electrons can be induced by a careful selection of the anode dimensions and voltage applied, resulting in proportional (multiplication) or ionisation (no multiplication) mode. In the multiplication process electron/ion couples are produced: the electrons are collected on the anode whereas the ions drift with a lower velocity towards the cathode.
Both of these contribute to the final signal, with the size of each component depending on the detector gain and the radial position of the pair creation.\\
More details on the detector working principle can be found in Ref.~\cite{Bouet:2020lbp,Bouet:2022kav} and references therein. In this section details are given on the SPS and CPC setups operated at LP2I Bordeaux and used to obtain the results presented in this paper.

\subsection{Spherical Proportional Counter}
\label{labelSPC}
\begin{figure} [t]
\centering
\subfigure[\label{fig:det1}]{\includegraphics[height=8
cm]{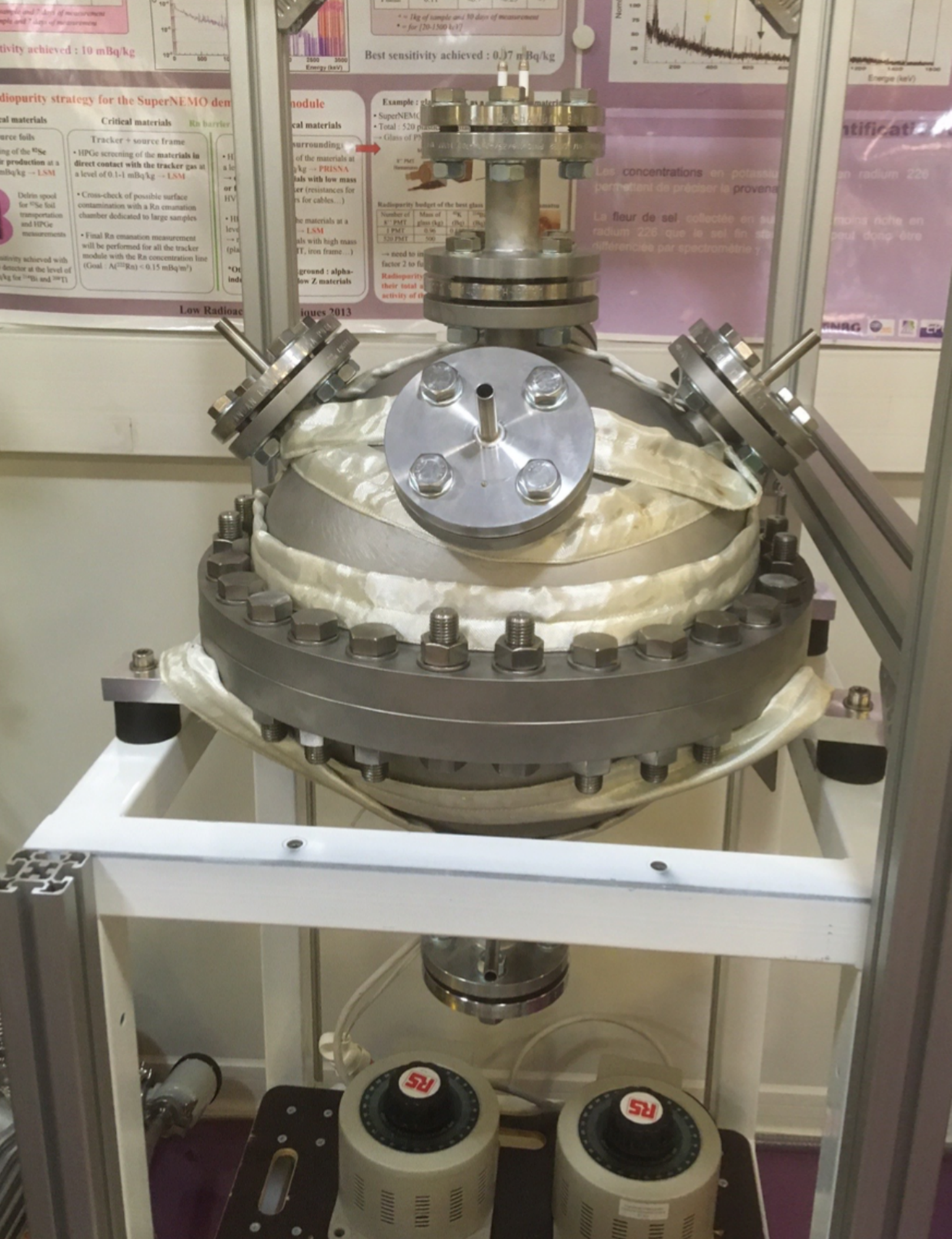}}
\subfigure[\label{fig:det2}]{\includegraphics[height=8
cm]{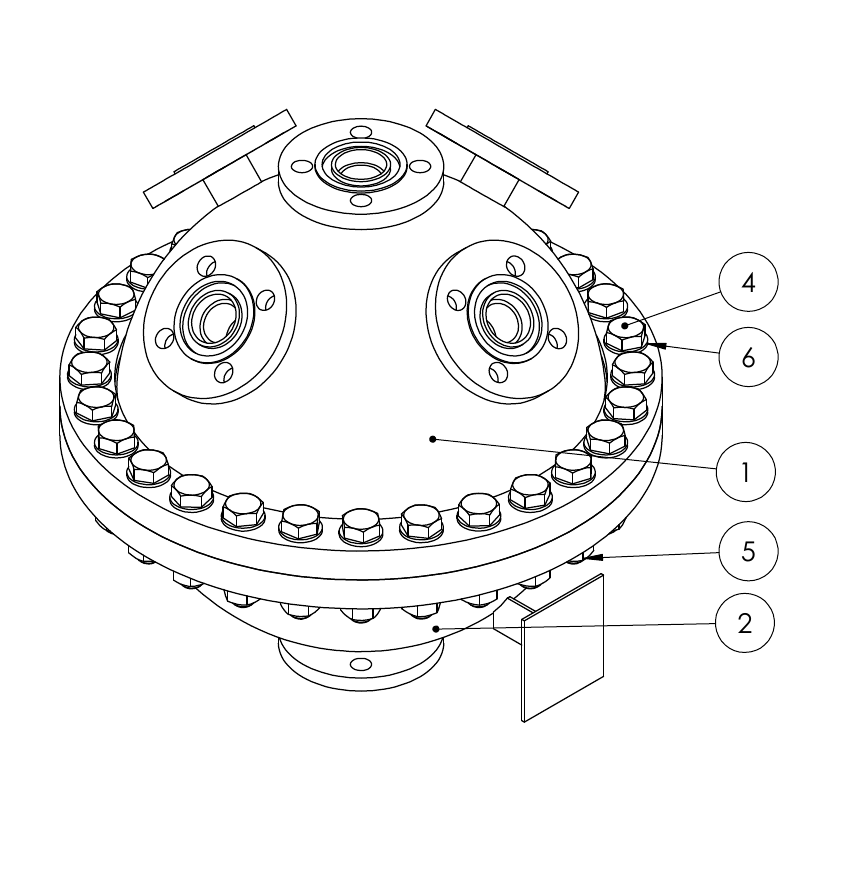}} \caption{{\it \subref{fig:det1}~Actual realisation and \subref{fig:det2}~mechanical drawing of the R2D2 detector. The numbers on the drawing correspond to different items used by the manufacturer.}}
\label{fig:1}
\end{figure}

The SPC built at LP2I Bordeaux in the framework of the R2D2 project consists of a 20~cm radius sphere which could contain about 8.5~kg of xenon at a pressure of 40~bar. Since there are no radiopurity constraints in the present R\&D phase, the material was chosen in a way to minimize the cost and facilitate the detector construction. The sphere was built in stainless steel by the RAVANAT company who took care of the certification needed to operate such a detector up to 40~bar. In Fig~\ref{fig:1} the conceptual plan and the actual detector are shown.\\
On the top hemisphere there are five different flanges: four of them are used for gas handling (pumping, filling, recirculating, and measuring vacuum and pressure). The central one includes a high-voltage (HV) feedthrough used to supply voltage to the central anode and also for the readout signal. Such a flange also serves as support for the rod holding the spherical anode in the middle of the detector. The flange in the bottom hemisphere is instead used to slide a  $^{210}$Po source, which emits 5.3~MeV $\alpha$ particles, into the detector to a radial distance of 20~cm from the center.\\
The central sensor is the core of the detector and the most delicate part: it has to withstand high-voltage up to several kV without discharging on the grounded rod, and minimize the noise as the same wire is used for both providing the positive HV on the sensor and extracting the signal. The sensor is built based on the collective expertise and R\&D of the collaboration~\cite{Katsioulas:2018pyh} and exploited today by the NEWS- G collaboration.\\
The signal is separated from the HV and amplified thanks to a dedicated electronics box built at LP2I Bordeaux which includes a filter and a charge amplifier. Details on the custom electronics, built in the framework of the OWEN (Optimal Waveform recognition Electronic Node) project~\cite{OWEN}, can be found in Ref.~\cite{Bouet:2022kav}. Stability of the HV is critical to reduce the noise, therefore a highly stable HV power supply is required despite the HV filter employed to smooth HV variation. Several power supplies were tested and the one selected is an iSEG EHS 8060p with a voltage ripple at the level of 3~mV. The signal is finally sampled and registered through the data acquisition (DAQ) made of a CALI card read by the SAMBA acquisition software~\cite{Armengaud:2017rzu}.\\

\subsection{Cylindrical Proportional Counter}

\begin{figure} [t]
\centering
\subfigure[\label{fig:det3}]{\includegraphics[height=8
cm]{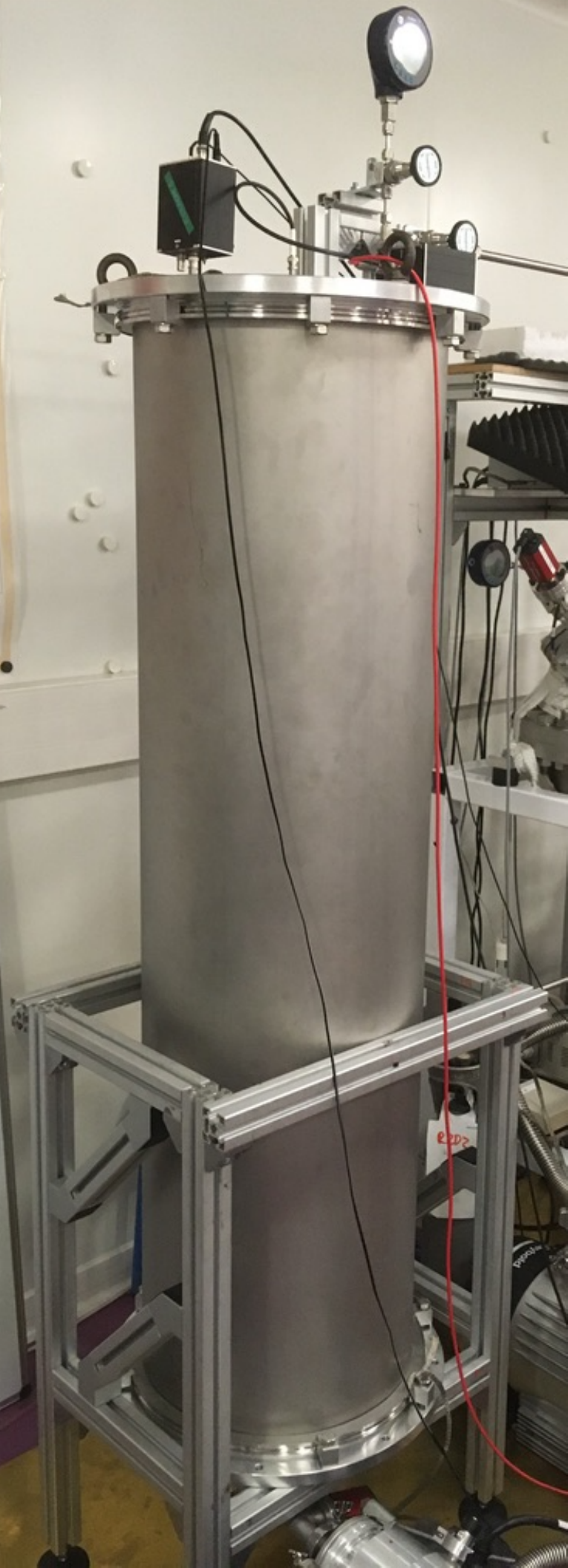}}
\subfigure[\label{fig:det4}]{\includegraphics[height=8
cm]{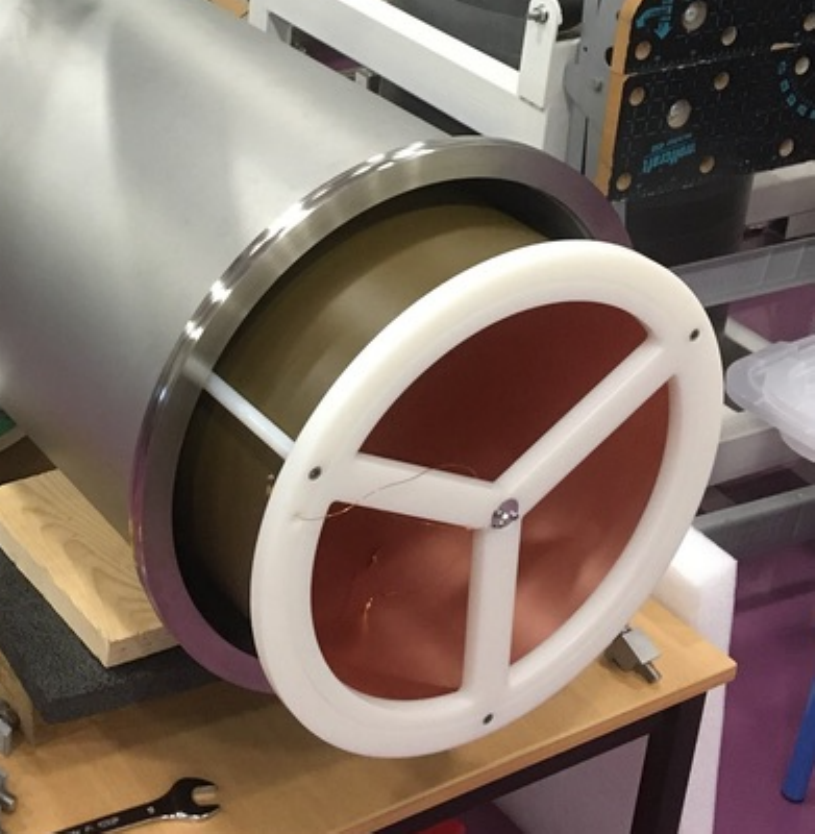}} \caption{{\it \subref{fig:det3}~Final installation of the CPC and \subref{fig:det4} inner structure.}}
\label{fig:2}
\end{figure}

\begin{figure} [t]
\centering
\includegraphics[width=0.75\textwidth]{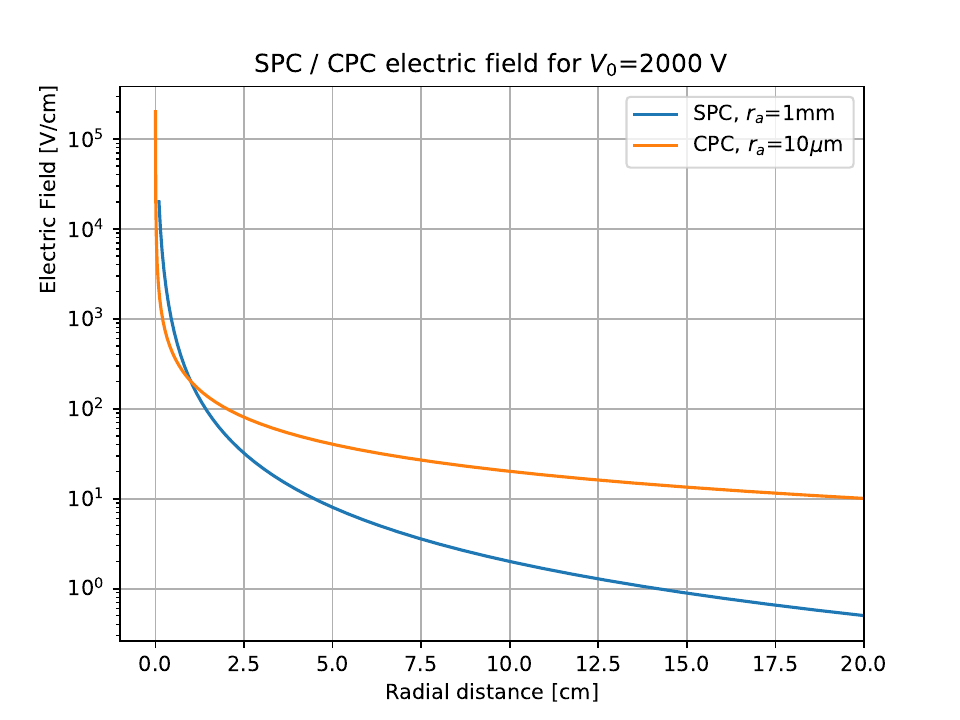}
\caption{{\it Comparison of the electric field as a function of the radial distance for a SPC with a central anode of 1~mm radius (blue line) and the one of a CPC with a central wire of 10~$\mu$m (orange line). In both cases the $\Delta$V between anode and cathode has been set to 2000~V.}}
\label{fig:2bis}
\end{figure}

The CPC was built and designed at SUBATECH and is composed of two parts. The outer part is a stainless steel cylinder 1.5~m high with a radius of 20~cm which serves as gas reservoir. The inner part, the CPC unit itself, is a cylinder 1~m high with a radius of 17~cm which is inserted into the tank. Although the assembly can be done horizontally, the tank is foreseen to be operated vertically. No pressure certification exists and the tank is not designed to be operated at pressures above the atmospheric one.
The upper flange of the tank is equipped with two feedthroughs (HV and signal reading) and a gas inlet for filling the detector, whereas the bottom one is equipped with a gas outlet for recirculation and another outlet for pumping the detector before filling. 
The CPC structure is composed of 2 nylon end-caps (3 branches) and 
3 fiberglass support columns. Such a skeleton holds a rolled sheet of GI180 (300 $\mu$m thick) which acts as a cathode, having the inner face covered with 
20 $\mu$m of copper. The central anode is a tungsten wire with a diameter of 20~$\mu$m, attached to the upper crosspiece and stretched at its lower end by a 
mass of 7~g to maintain the necessary tension. Such a layout defines an active CPC volume of about 90 litres. A view of the detector and its inner structure 
can be seen in Fig.~\ref{fig:2}.\\
Electronics, DAQ and signal processing are identical to those used for the SPC. However, an important feature differentiates the CPC from the SPC: the central wire is grounded and a negative HV is applied to the cathode. 
Such a reversed-bias configuration has the advantage of decoupling the signal from the HV, making the noise independent of the HV applied. 
Moreover, the electric field decreases radially as 1/R, resulting in a much stronger field near the cathode. A comparison between the field of a SPC with a central anode of 1~mm radius and the one of a CPC with a central wire of 10~$\mu$m radius can be seen in Fig.~\ref{fig:2bis}. Because of the different electric field, the CPC signals differ slightly from the SPC ones. The stronger field of the CPC far from the anode leads to a shorter drift time and lower diffusion effect, therefore shorter rise time and lower attachment effects are expected. 

\section{Laboratory setup}
\begin{figure} [t]
\centering
\subfigure[\label{fig:det5}]{\includegraphics[height=8
cm]{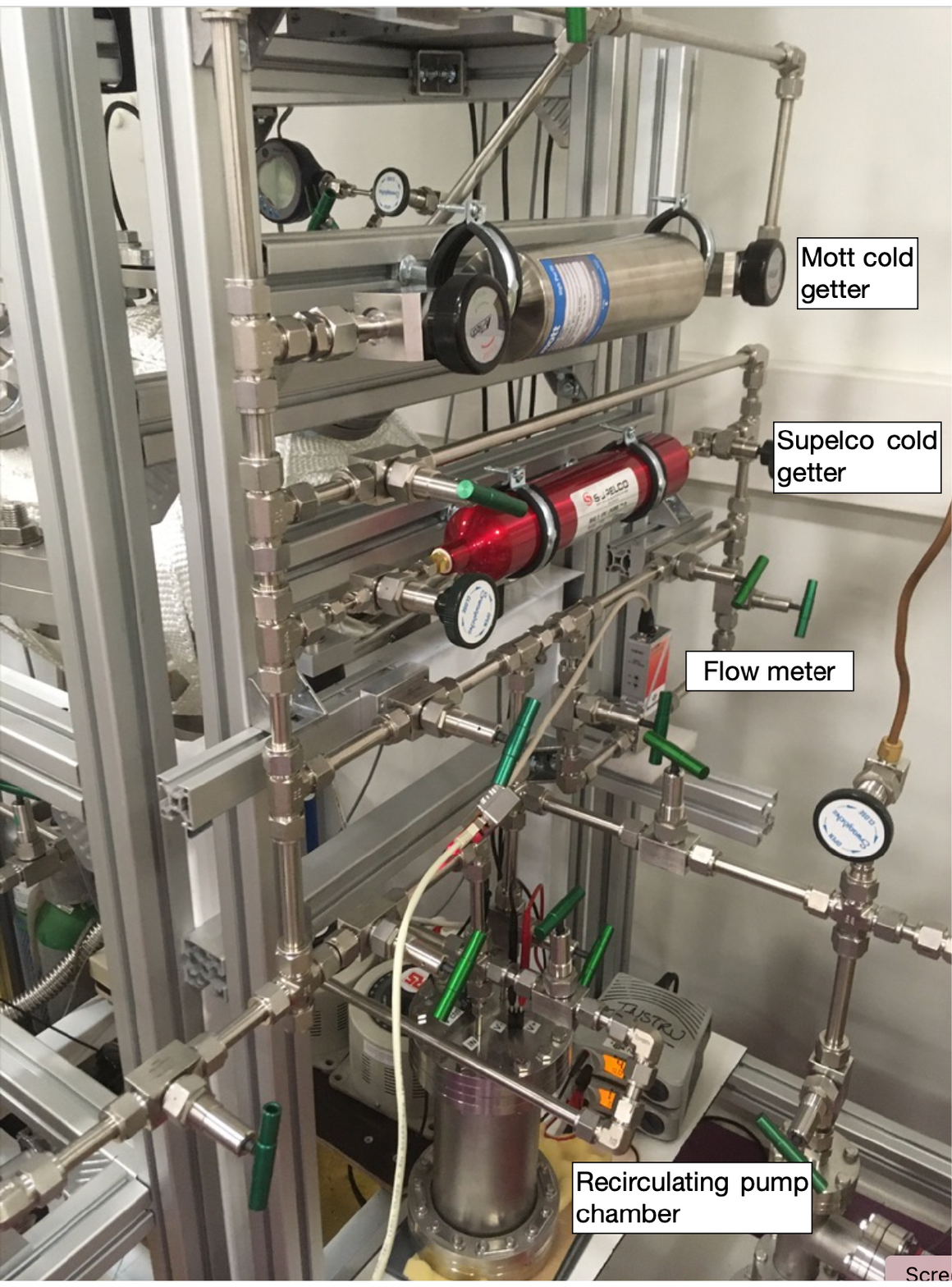}}
\subfigure[\label{fig:det6}]{\includegraphics[height=8
cm]{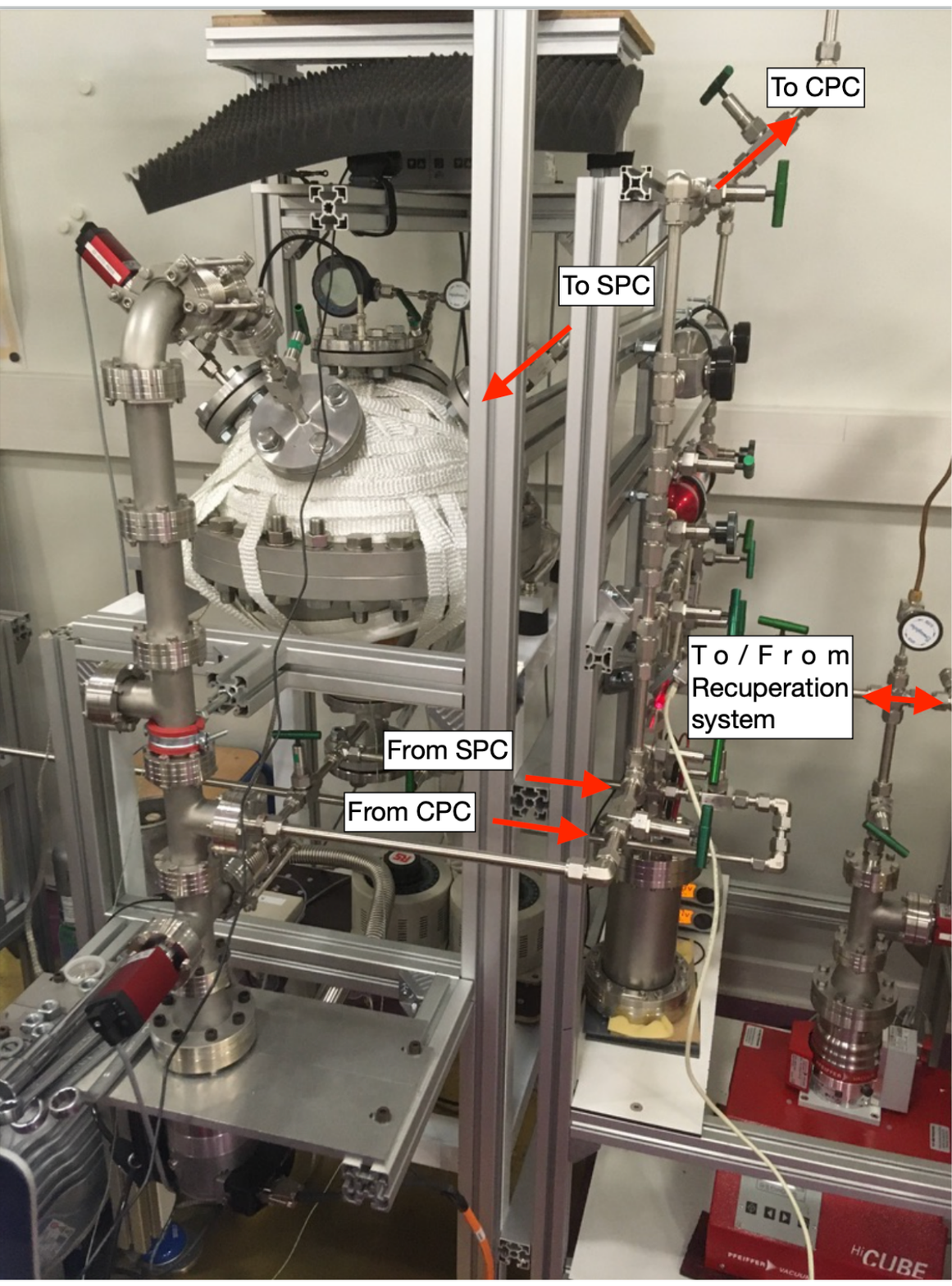}} \caption{{\it \subref{fig:det3}~Zoom of the recirculation system showing the pump and the two cold getters and \subref{fig:det4} full system view with indication of the different gas flows.}}
\label{fig:3}
\end{figure}

In order to reduce electronegative impurities such as oxygen, which result in a  loss of signal due to electrons attachment during the drift, a recirculation system is used to purify the gas. Moreover, due to the high cost of xenon, it is important to recover it after use.  A combined setup, conceived at CPPM, was developed at LP2I Bordeaux and used for both the SPC and the CPC (see Fig~\ref{fig:3}).\\
The recirculation system consists of a pump, enclosed in a dedicated sealed chamber to avoid any possible leak of xenon, which forces the gas through two cold getters: a Supelco cartridge followed by a Mott one. The first getter is less powerful in terms of purification, however a prior cleaning helps to preserve the Mott cartridge which can achieve sub-ppb contamination of oxygen. A flowmeter from the Bronkhorst company controls the gas flow through the purification cartridges. The pump velocity is set in order to have a stable flow at about 2 litres per minute. The system can be upgraded with the addition of a hot getter, further improving the gas quality. Such a modification could become essential for operation at pressure higher than the atmospheric one. Another possible future upgrade to further reduce the electronegative impurities in the gas could be the replacement of the commercial pump in use with a magnetically-driven piston pump, which was indeed developed for clean applications~\cite{LePort:2011hy}.\\
The recuperation system is based on cryogenic pumping. A one-gallon bottle is connected to the system and held in a dewar which can be filled with liquid nitrogen. When the bottle is cooled, it collects the xenon because of the lower pressure created. When equilibrium is reached, the remaining pressure in the detector is at the level of few mbars, resulting in minimal gas loss if the detector has to be opened for maintenance.\\

\section{Updated argon measurements}
\begin{figure} [t]
\centering
\includegraphics[width=0.75\textwidth]{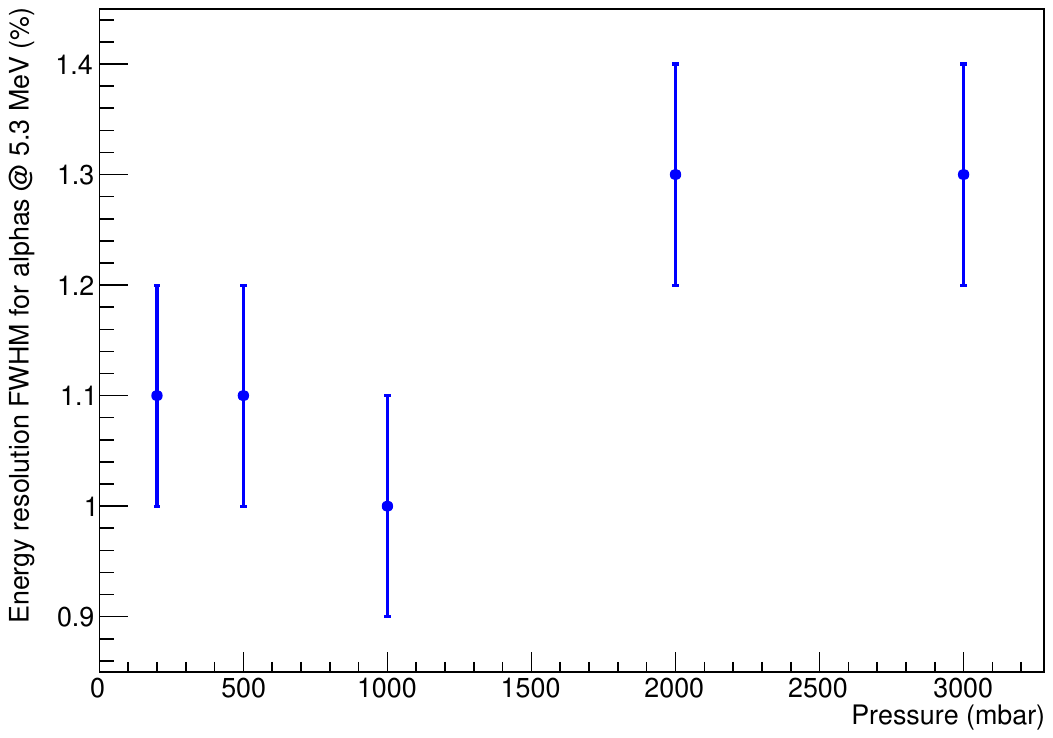}
\caption{{\it Energy resolution FWHM obtained for 5.3~MeV $\alpha$ signal in ArP2 with a SPC operating in proportional mode between 200~mbar and 3~bar.}}
\label{fig:4}
\end{figure}

\subsection{Measurements with the SPC}

Earlier ArP2 measurements, performed with the SPC at pressures between 200~mbar and 1.1~bar, showed that the energy resolution, approaching 1\% FWHM, was independent of the track length~\cite{Bouet:2020lbp}. The 1.1 bar limit stems from the experimental device, which cannot be certified for pressures greater than one atmosphere.\\
To overcome this obstacle, a new SPC, described in Sec.~\ref{labelSPC} and certified to be operated up to 40~bar, was commissioned at LP2I Bordeaux to enlarge the pressure scan range and validate the detector behaviour at higher pressure. As summarized in Fig.~\ref{fig:4}, the performed pressure scan indicates that 1.3\% resolution can be obtained up to 3~bar in proportional mode. This last restriction in pressure was caused by the voltage that the detector could withstand. Indeed, at 3~bar, the HV applied on the sensor reached 3900~V, and the small distance between the central anode and the supporting grounded rod started to be a critical issue: small discharges were seen impacting the gain stability over time and degrading the resolution.\\

In order to circumvent the voltage limit of the SPC, the detector was operated in ionization mode ({\it i.e.} at lower HV without an electron avalanche near the anode). In order to benefit from a high-drifting electric field at long distances from the anode, a larger central anode was needed. An anode with 3~mm radius (instead of the original 1~mm radius) was installed, allowing data taking at 1~bar with an HV of 700~V (instead of the 1900~V applied to the small sensor of 1~mm radius in proportional mode).  
In these operating conditions, the integral spread was approximately 2.5~ADU (DAQ Digital Units), the limit of the setup, the same width obtained using a perfect generator signal as input. The $\alpha$ particle signal, however, has an integral of $\sim$70~ADU, corresponding to a FWHM resolution of around 8\%. The $\alpha$ particle signal integral was independent of the HV applied (between 700~V and 1900~V) and the resolution is thus limited by baseline fluctuations increasing when the HV applied to the anode increases.\\

\subsection{Measurements with the CPC}
\begin{figure}[t]
\centering
\subfigure[\label{fig:5-1}]{\includegraphics[width=0.45\textwidth]{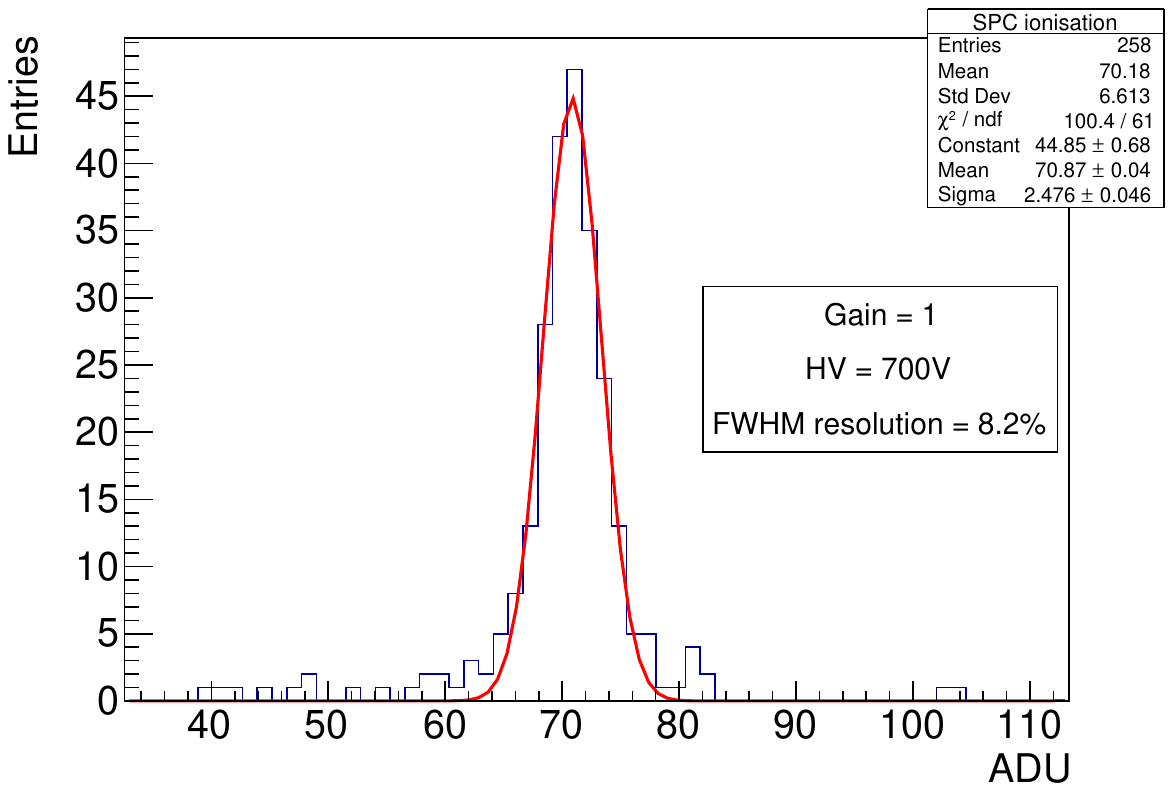}}
\subfigure[\label{fig:5-2}]{\includegraphics[width=0.45\textwidth]{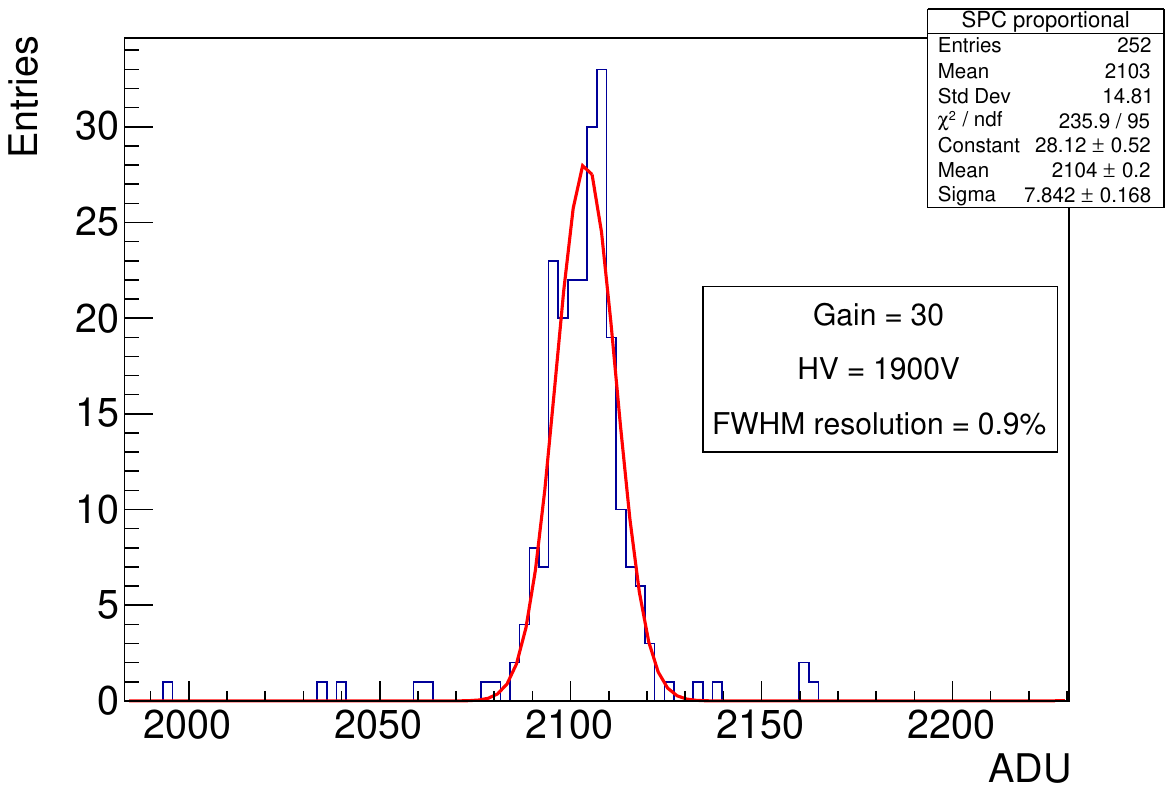}}
\subfigure[\label{fig:5-3}]{\includegraphics[width=0.45\textwidth]{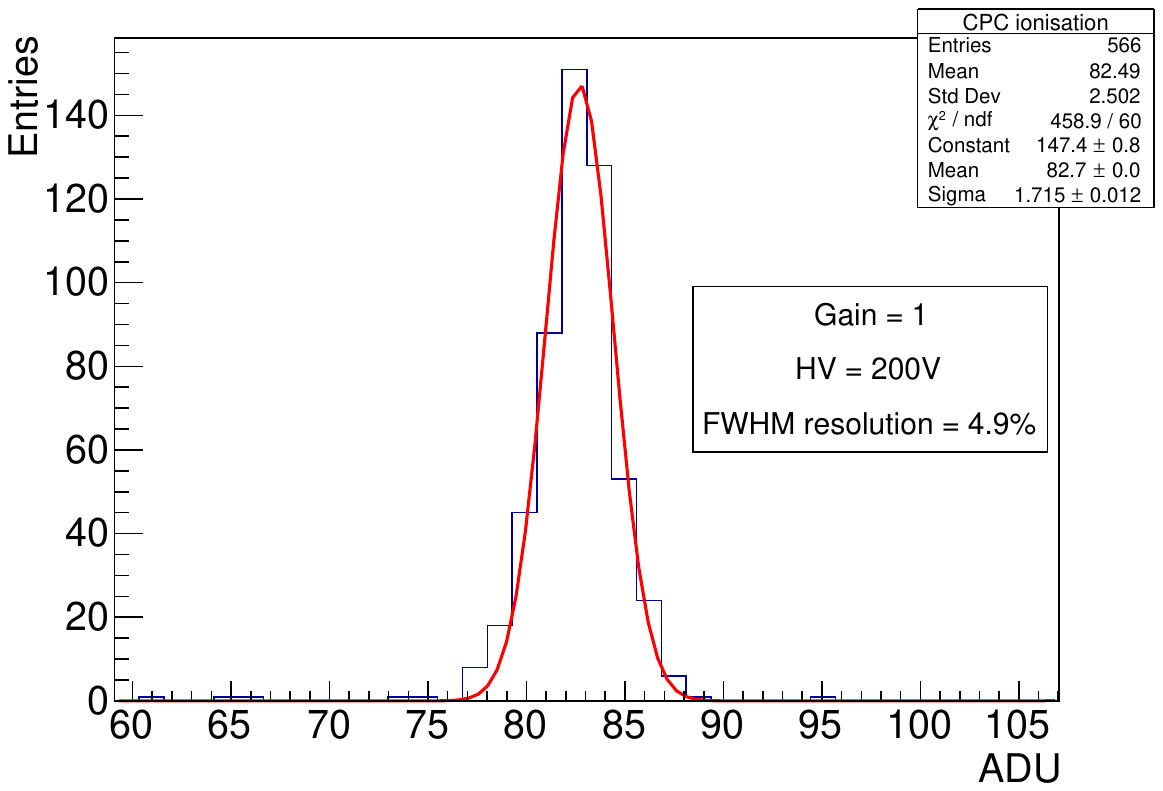}}
\subfigure[\label{fig:5-4}]{\includegraphics[width=0.45\textwidth]{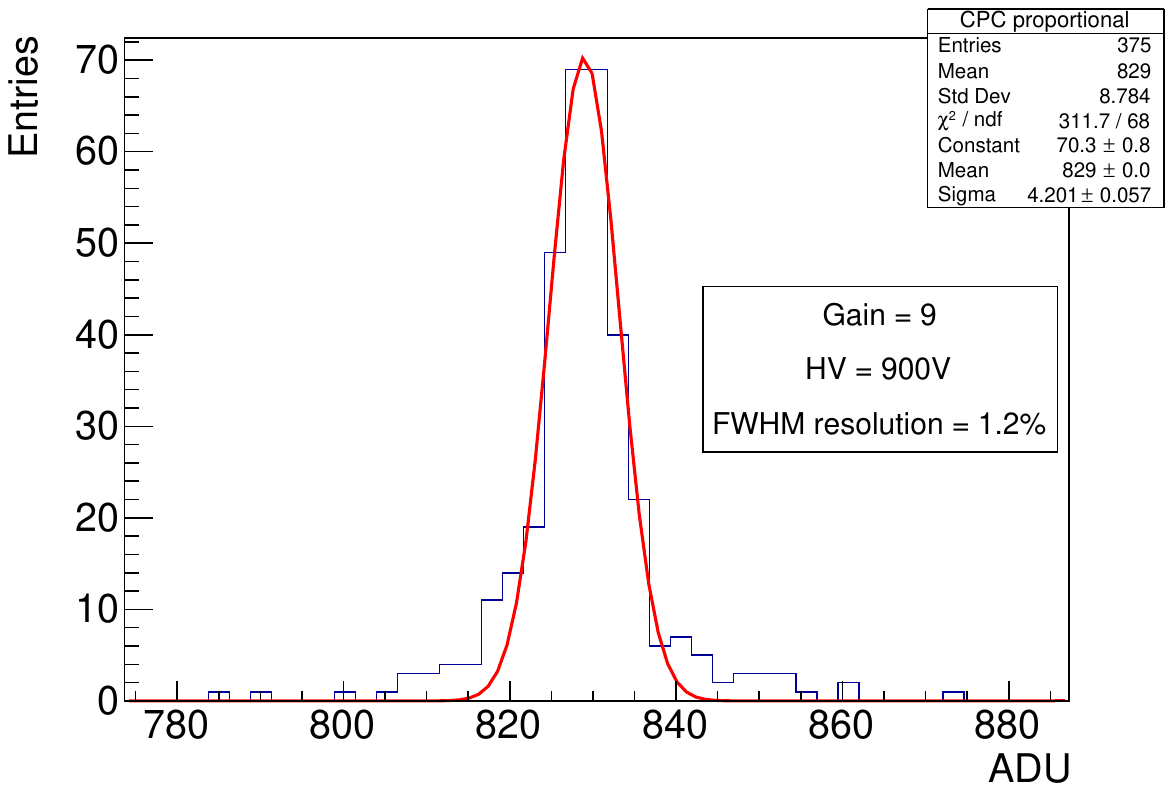}}

\caption{{\it Best results for energy resolution in ArP2 at 1~bar for \subref{fig:5-1} SPC in ionisation mode with 3~mm radius anode, \subref{fig:5-2} SPC in proportional mode with 1~mm radius anode, \subref{fig:5-3} CPC in ionisation mode, and \subref{fig:5-3} CPC in proportional mode.}}
\label{fig:5}
\end{figure}

As previously mentioned, the existing CPC can not be operated at a pressure higher than 1~bar. It was therefore tested filled with ArP2 at 1~bar confirming that the separation of signal and HV results in a lower noise level with respect to the SPC. Applying a HV of 900~V, a resolution of 1.2\% was obtained in proportional mode, similar to that obtained with the SPC but with a much lower HV, as expected ({\it i.e.} 900~V as opposed to 1900~V for a similar gain). A test was performed in ionisation mode as well, obtaining a resolution of 5\%. Such a resolution was again limited by the baseline noise, which remained contained at around 1.7~ADU thanks to the grounded anode.\\

\subsection{CPC / SPC comparison in argon}

The best results obtained at 1~bar with ArP2 for the different setups are shown graphically in Fig.~\ref{fig:5}, where the gain and the HV are also stated. The only selection cut applied on data is a cut on the signal risetime: events with a small risetime correspond to $\alpha$ particles hitting the cathode and releasing only a fraction of their energy inside the gas. Such a cut removes the low energy tail in the energy distribution ({\it i.e.} the integral of the recorded signal).

To continue the study at higher pressure with the SPC a sensor designed to be operated at a HV up to about 10 kV has to be specifically designed. Such a study is ongoing thanks to a collaboration with AXON company and results are expected soon.\\  
The only limitation currently identified in the CPC setup is the pressure certification: as soon as this is overcome the same study can be performed up to 40~bar.   
In view of the published literature, the HV will not constitute a prohibitive problem for the CPC up to 20~kV, at least as long as the cathode is sufficiently separated from the outer cylinder in order to avoid possible voltage breakdowns.\\

\section{First xenon measurements}
\begin{figure} [t]
\centering
\subfigure[\label{fig:det61}]{\includegraphics[height=4.1
cm]{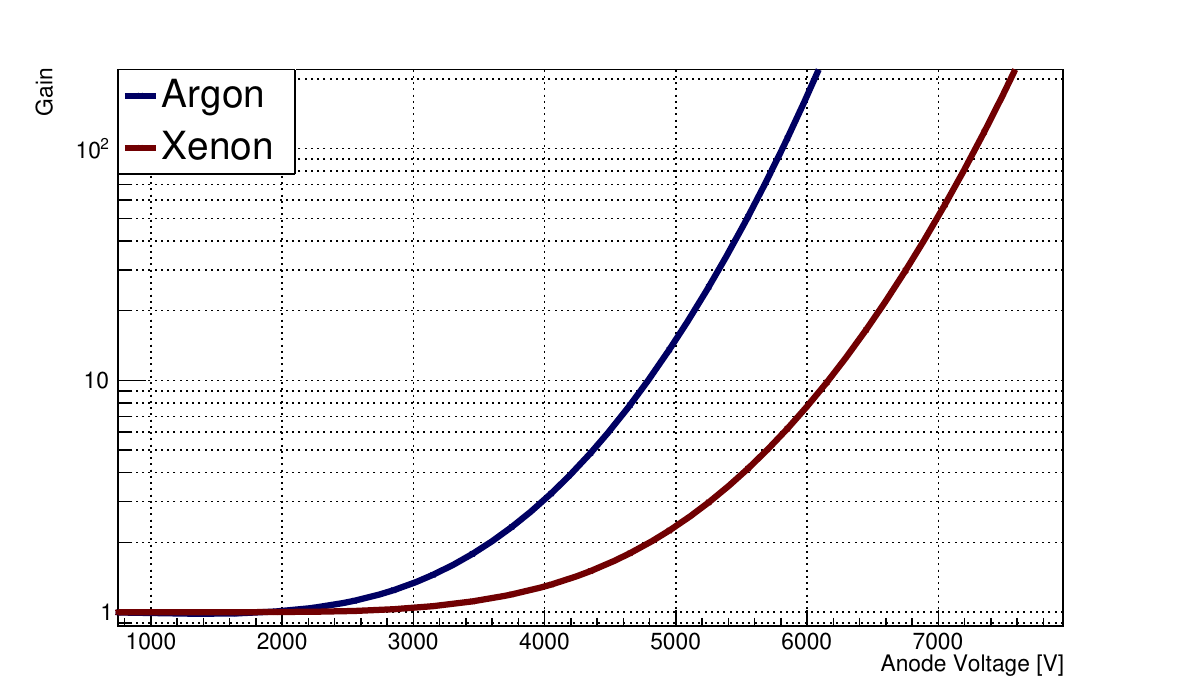}}
\subfigure[\label{fig:det62}]{\includegraphics[height=4.1
cm]{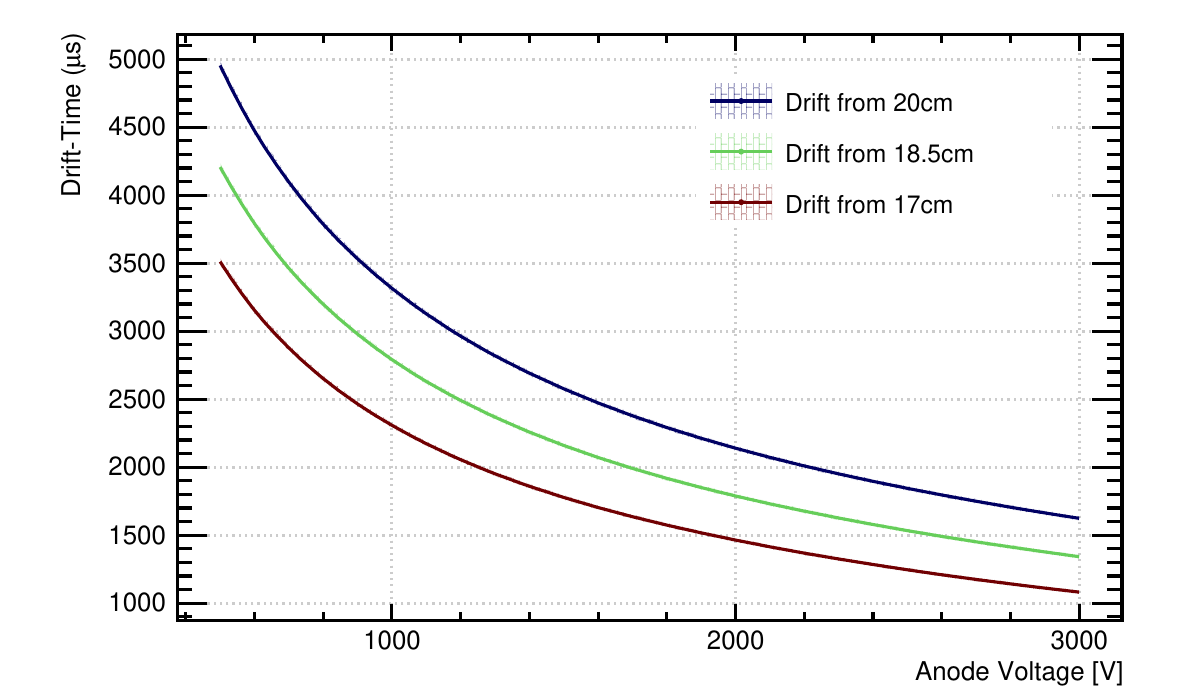}} \caption{{\it \subref{fig:det61}~Gain Vs. anode voltage for argon (blue) and xenon (red) at 1~bar for a SPC with a 3~mm radius anode. \subref{fig:det62}~Electrons drift time for different radial position in xenon at 1~bar for a SPC with a 3~mm radius anode as a function of the anode voltage.}}
\label{fig:6}
\end{figure}

\subsection{Measurements with the SPC}
The major challenge in moving from argon to xenon is given by the gas purification requirements:  for the same pressure and HV, the electron drift time is about a factor of 10 slower in xenon with respect to argon. This means that in xenon the effect of electronegative attachment gets higher, thus reducing the total charge collected, and therefore degrading the energy resolution. For this reason, the 1~mm anode radius was deemed too small to guarantee an efficient collection of electrons created close to the cathode, and the 3~mm SPC anode, already tested in argon, was selected for the xenon measurements.\\
In parallel,  a full Monte-Carlo simulation based on Geant4~\cite{GEANT4:2002zbu}, COMSOL~\cite{comsol} and Garfield++~\cite{garfield} was set up to validate the detector understanding (more details on the simulations can be found in Ref.~\cite{Bouet:2020lbp}). Particular attention was paid to the drift time and the gain as a function of the HV and the nature of the gas, Ar or Xe, at 1 bar. The results presented in Fig.~\ref{fig:6} indicate that only the ionisation mode is viable for a SPC in xenon without applying very high voltages: for a gain of 10 at 1~bar the required HV is already around~6 kV.\\
An initial test at 250~mbar was carried out with a HV scan from 800~V to 1400~V. The upper limit in HV comes both from the fact that the baseline noise depends on this setting, and that the baseline fluctuation should not exceed the relevant threshold needed to trigger on actual source events. The optimal working point was obtained at 1300~V with an integral of 118~ADU and a $\sigma$ of 1.9~ADU corresponding to a resolution of 3.8\%.\\
Another scan, at increased pressure of 900~mbar, was completed between 1300 and 2200~V. In this case the resolution was rather stable up to 2000~V with an integral of about 85~ADU and a $\sigma$ of about 2.5 ADU giving a resolution of 7.2\%. The gas pressure was successively increased to 2~bar in order to check the stability of the obtained results at higher pressure. However, even after 48 hours recirculation it was hard to observe the signals due to the electrons attachment. Indeed, a dependence of the signal integral as a function of the signal risetime was observed, which is typically absent in the case of no electron attachment, confirming that the limiting factor comes from the gas purification efficiency.\\

\subsection{Measurements with the CPC}

\begin{figure} [t]
\centering
\includegraphics[width=0.75\textwidth]{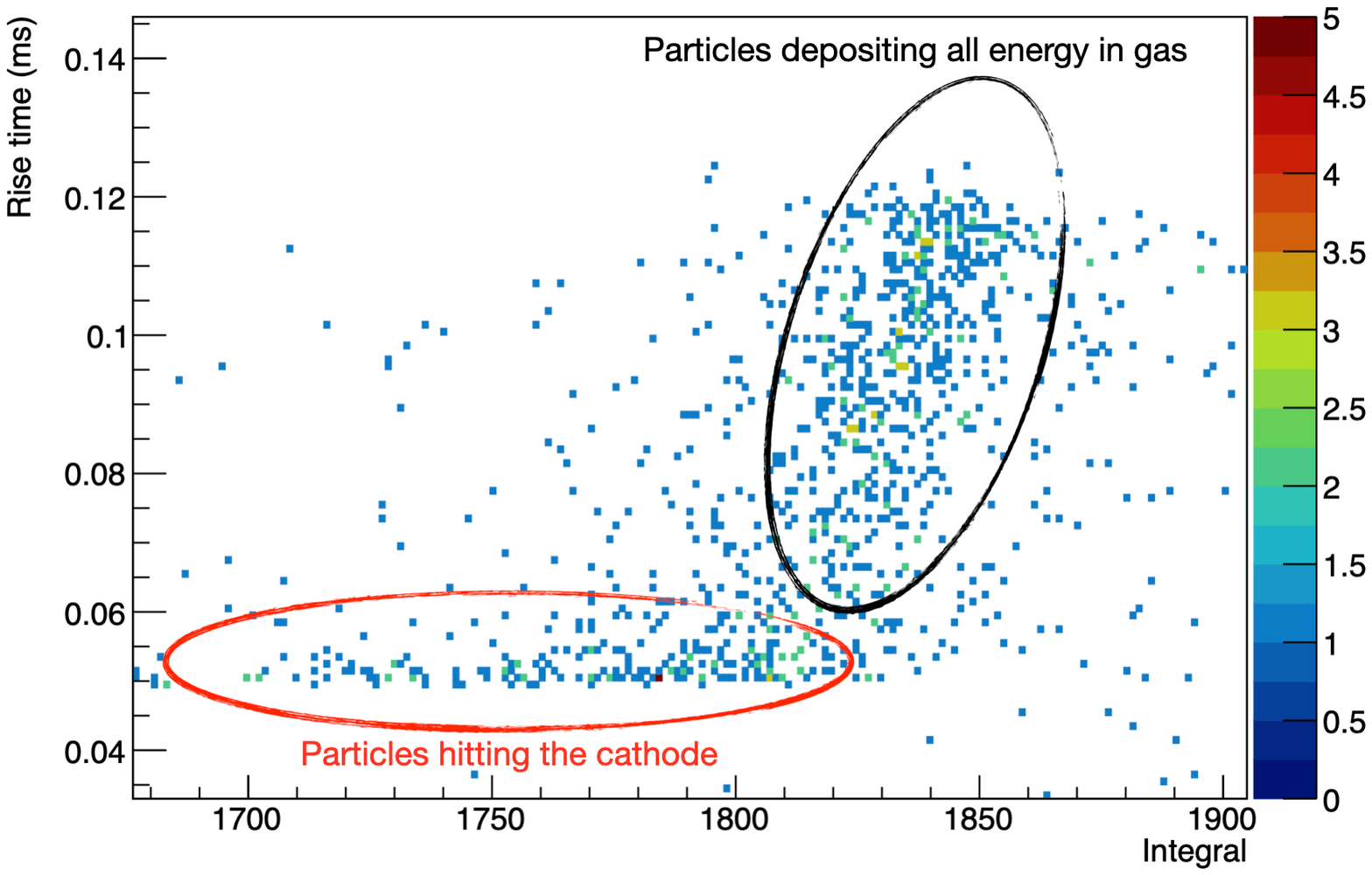}
\caption{{\it Risetime Vs integral for CPC operated in xenon at 500 mbar and 900~V.}}
\label{fig:7}
\end{figure}

\begin{figure} [t]
\centering
\subfigure[\label{fig:det81}]{\includegraphics[height=5
cm]{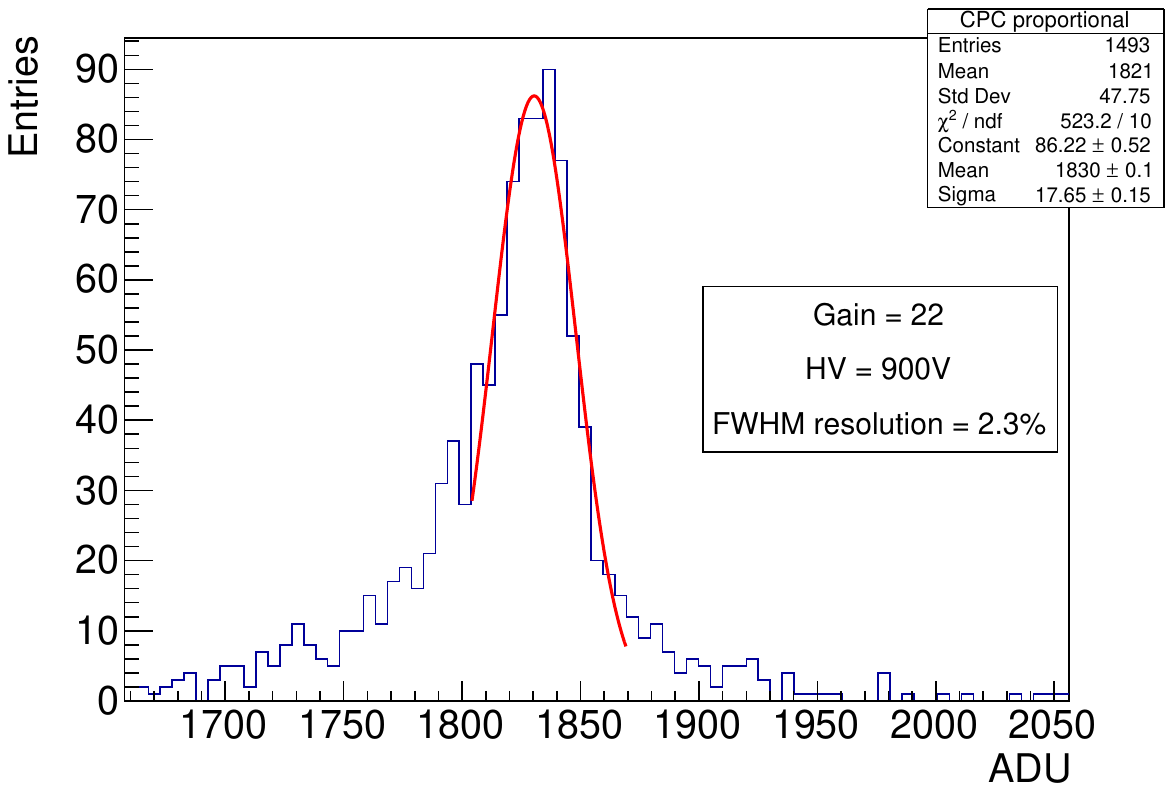}}
\subfigure[\label{fig:det82}]{\includegraphics[height=5
cm]{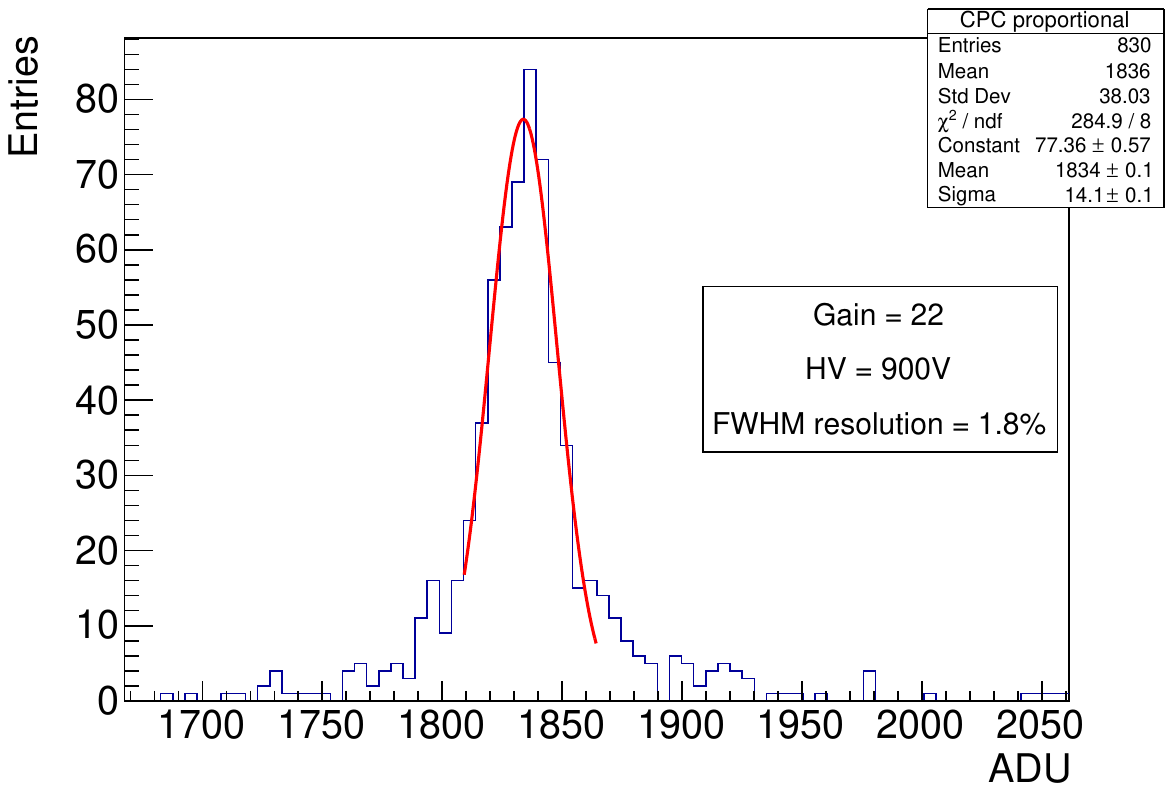}} \caption{{\it Resolution obtained for a CPC in xenon at 500 mbar and 900~V with \subref{fig:det81} no cut on risetime, and \subref{fig:det82} a cut on risetime at 0.07~ms.}}
\label{fig:8}
\end{figure}

The advantage of the CPC is the option to work in proportional mode. This geometry was first tested at 500 mbar. After 24 hours of gas recirculation the attachment was still visible in the slope of the bi-plot risetime versus integral, as shown in Fig.~\ref{fig:7}.
For a sufficiently pure gas, no attachment due to electronegative impurities is expected. The reconstructed charge ({\it i.e}. the signal integral) therefore becomes independent of the radial position of the deposited energy. On the contrary, in the presence of electronegative impurities, the primary electrons emitted farther from the anode have a higher probability of being captured, resulting in a lower induced charge. This behavior is clearly visible on the plot of Fig.~\ref{fig:7}: particles going towards the cathode, with a larger risetime, are less affected by the electronegative attachment and a larger energy deposition is reconstructed.\\
With the obtained gas quality, a resolution of 2.3\% FWHM has been achieved: this can be improved to 1.8\% by selecting events with a risetime greater than 0.07~ms, rejecting $\alpha$ particles partially contained in the detector. This
is shown in Fig.~\ref{fig:8}. These results clearly show that the purity of the gas plays a major role. With a longer recirculation and the use of a hot getter, not available at the time of the data taking, further improvement is still expected.

\begin{figure} [b]
\centering
\subfigure[\label{fig:det91}]{\includegraphics[height=5
cm]{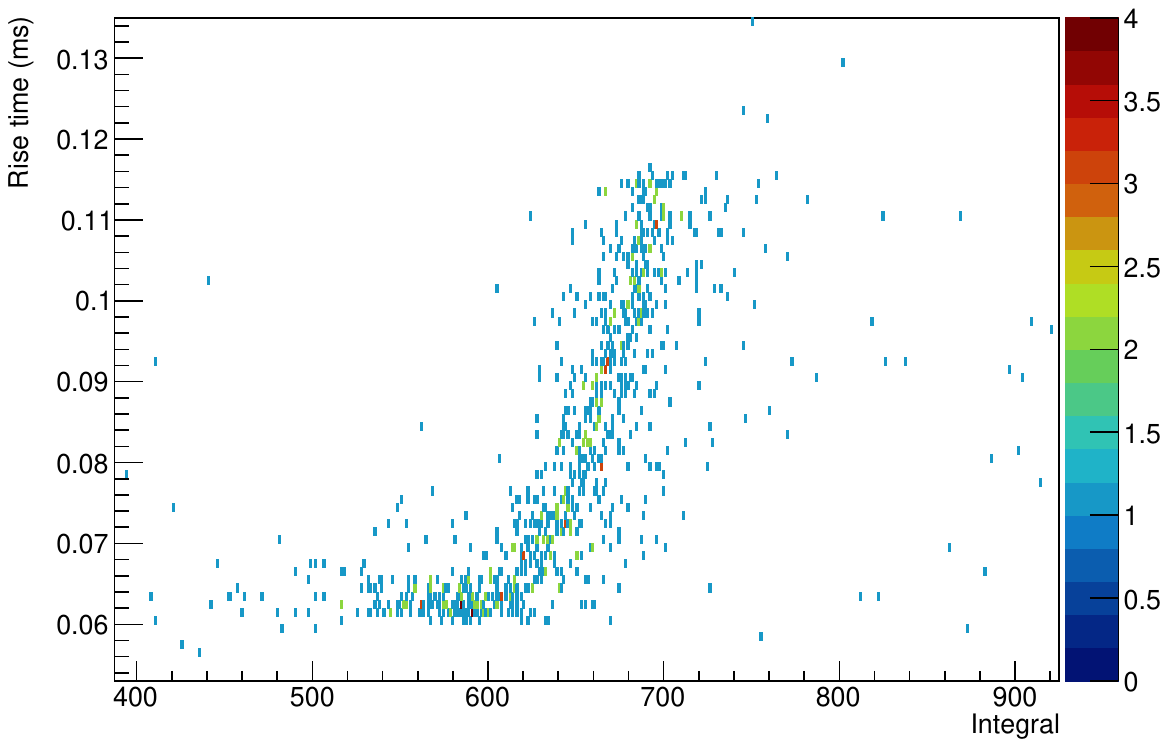}}
\subfigure[\label{fig:det92}]{\includegraphics[height=5
cm]{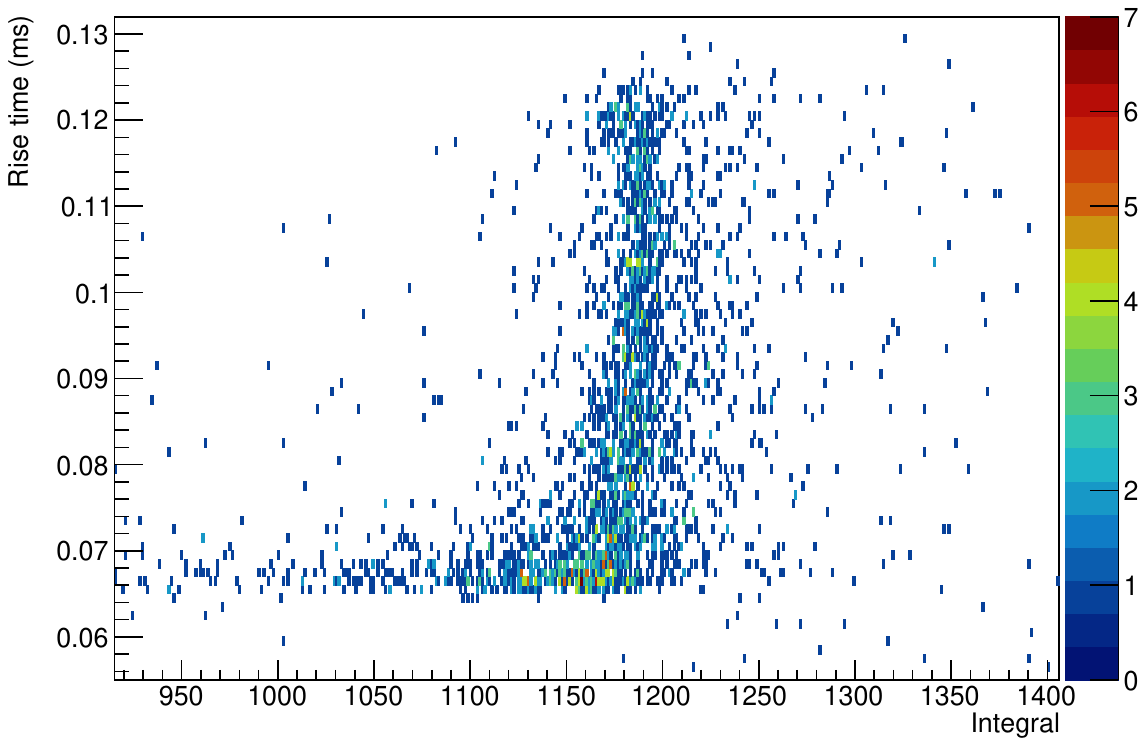}} \caption{{\it Risetime Vs signal integral ({\it i.e.} reconstructed charge) for CPC operated in xenon at 1 bar and 1200~V after \subref{fig:det91} 24 hours and \subref{fig:det92} 48 hours or recirculation.}}
\label{fig:9}
\end{figure}

The pressure was increased to 1~bar, the maximal pressure allowed by the present setup. At this pressure, primary ionisation track lengths are short (1.5~cm) but also purity-related effects become more prominent. Upon recirculation of the gas through the two purification cartridges, a continuous improvement in gas purity was observed, but stability in detector response was only achieved after approximately 48 hours. It manifested itself in a twofold effect. First, the signal integral becomes independent of the particle direction ({\it i.e.} on the risetime) since all electrons reach the anode independently of their production point. Second, the absolute signal integral increases since more electrons reach the anode with respect to the number of collected electrons in the presence of attachment. Both effects are highlighted by Fig.~\ref{fig:9} where the plot of the risetime versus the signal integral is shown after 24 and 48 hours of gas recirculation.

\begin{figure} [t]
\centering
\subfigure[\label{fig:det101}]{\includegraphics[height=5
cm]{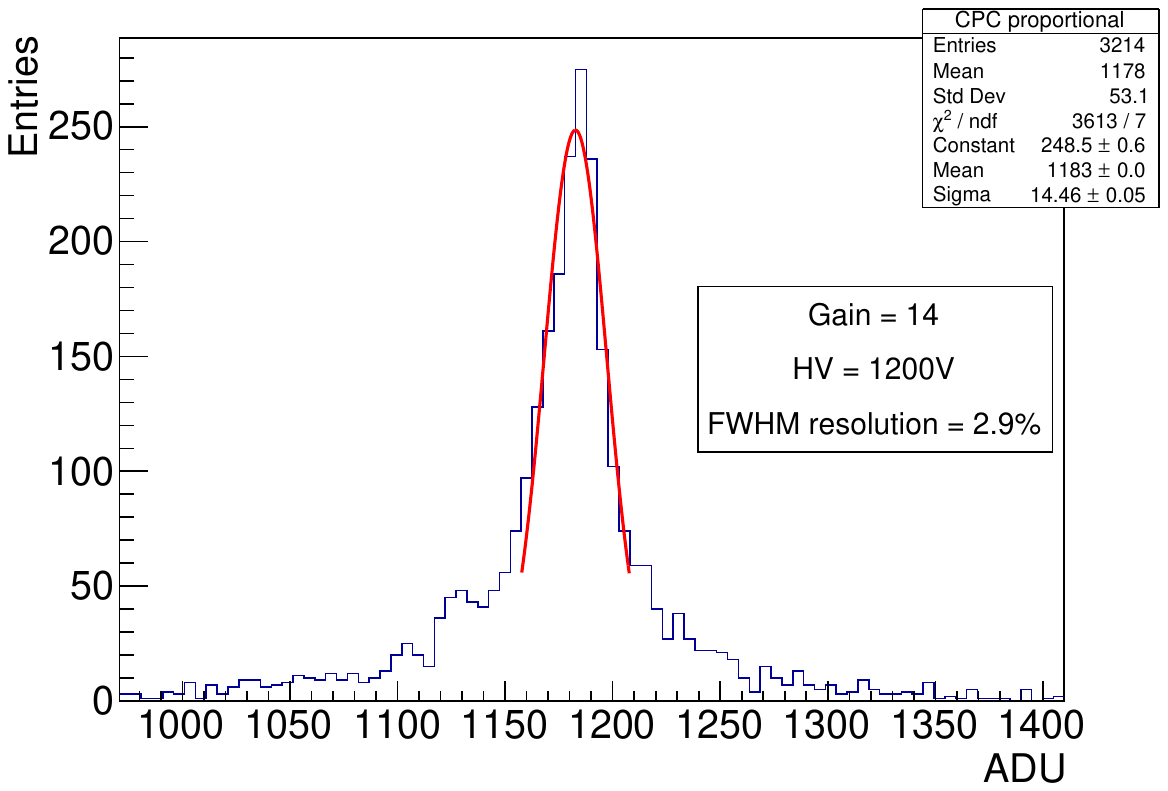}}
\subfigure[\label{fig:det102}]{\includegraphics[height=5
cm]{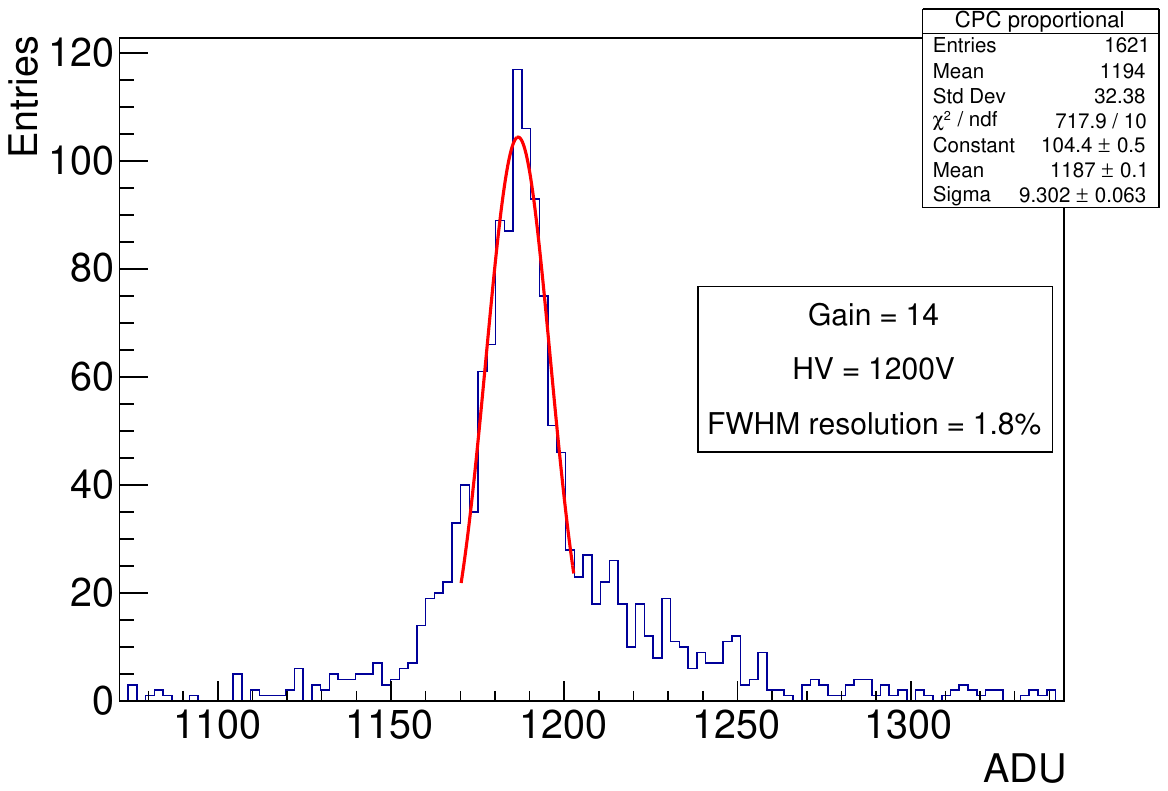}} \caption{{\it Resolution obtained for a CPC in xenon at 1 bar and 1200~V \subref{fig:det101} without any cut, and \subref{fig:det102} with a cut on risetime at 0.08 ms.}}
\label{fig:10}
\end{figure}

\begin{figure} [b]
\centering
\includegraphics[width=0.55\textwidth]{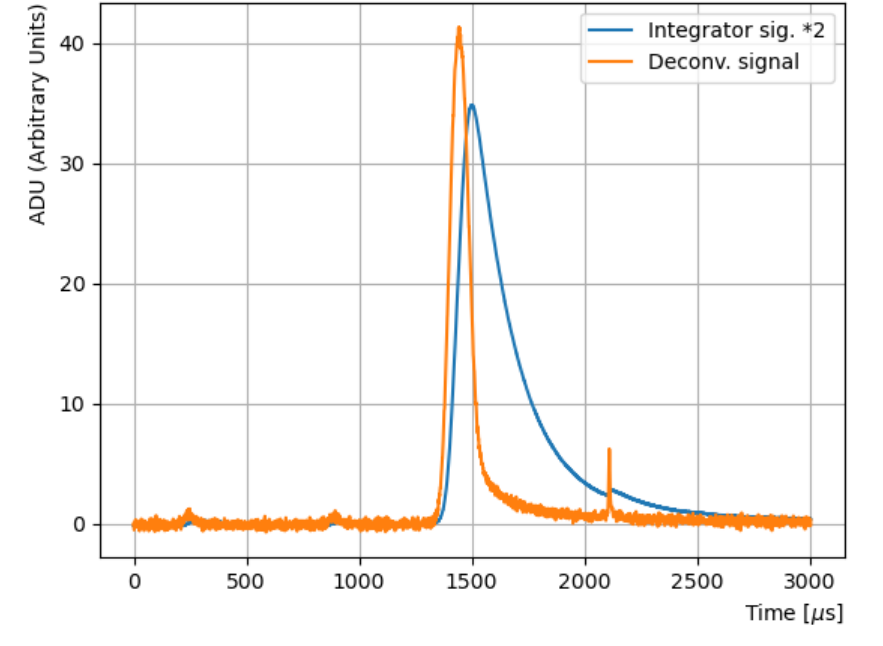}
\caption{{\it Waveform of one signal event in a CPC operated with xenon at 1 bar and 1200~V. The blue line represents the integrated signal whereas the orange one is the signal after the deconvolution from the preamplifier RC. The presence of a cosmic event on the tail of the signal is identified by the peak at about 2100~$\mu$s.}}
\label{fig:11}
\end{figure}

The resolution was computed after 48 hours of recirculation, once the gas reached the best purity allowed by the present setup. A resolution of 2.9\% was obtained without applying any selection cut.  A selection cut on the risetime to reject partially contained events, resulted in a resolution improvement to the level of 1.8\% as shown in Fig.~\ref{fig:10}.\\
In the end, by increasing the pressure from 500~mbar to 1~bar (and despite the loss in gain of 35\%), the resolution remained relatively identical. It was concluded that the obtained results were mainly driven by the gas purity, which could be improved with the addition of a hot getter in the recirculation circuit.\\ 
However, compared to the SCP results, an additional source of energy resolution degradation was introduced due to the larger volume and therefore the increased sensitivity of the CPC to cosmic muons. Indeed, given the small overburden of the detector location at LP2i Bordeaux, a significant cosmic muon flux traverses the larger volume of the CPC setup. A muon crossing 50~cm of active volume of xenon at 1~bar releases $\sim$0.3 MeV, a non-negligible contribution to the 5.3~MeV alpha particles. This explains the right-side tail of the reconstructed signal integral distributions. The presence of cosmic muons can be clearly identified looking at the signal waveforms: an example is shown in Fig.\ref{fig:11}.\\
To reject those events, a muon veto could be installed on the detector and an anti-coincidence system be implemented. Nonetheless, this background does not represent a real issue since the detector should be operated underground in the context of the $\beta\beta0\nu$ search, therefore avoiding such a waveform distortion.\\
A specific analysis was carried out looking for events with an additional peak on the tail indicating the presence of a cosmic muon induced pile-up and impacting the reconstructed energy. Such events were discarded and a pure sample was selected with an efficiency of about 26\%. An energy resolution of 1.4\% was obtained for the selected sample.

\begin{figure} [h]
\centering
\subfigure[\label{fig:ResSPC}]{\includegraphics[height=5.
cm]{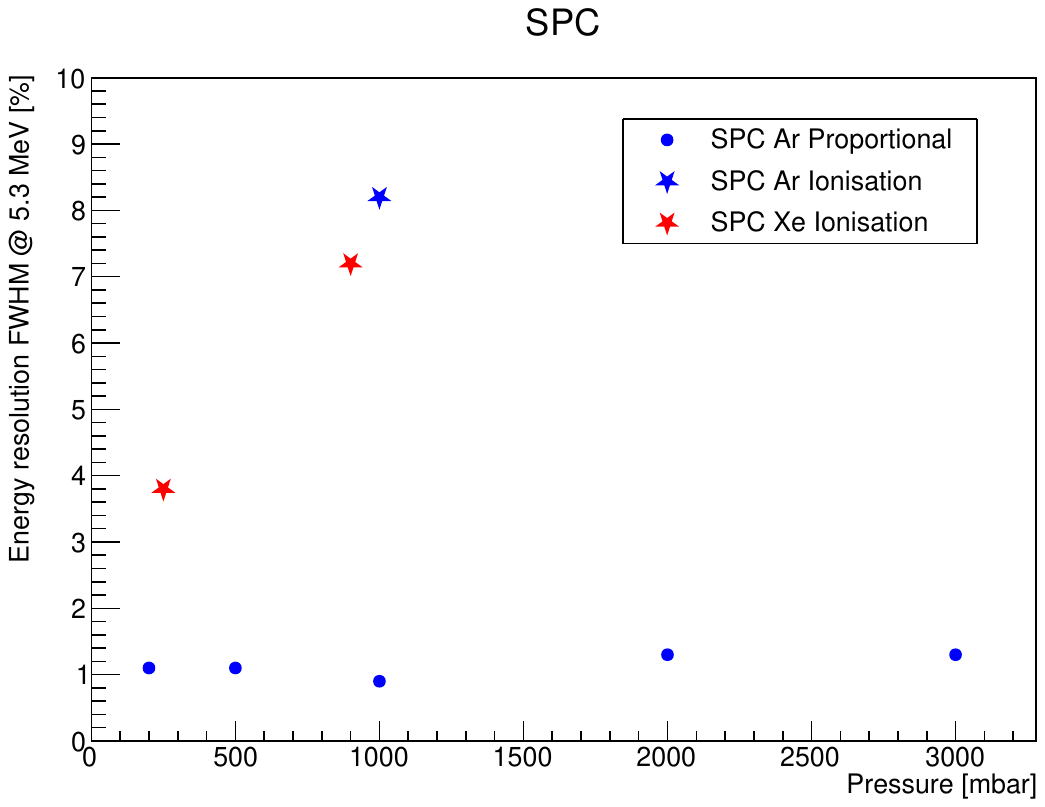}}
\subfigure[\label{fig:ResCPC}]{\includegraphics[height=5.
cm]{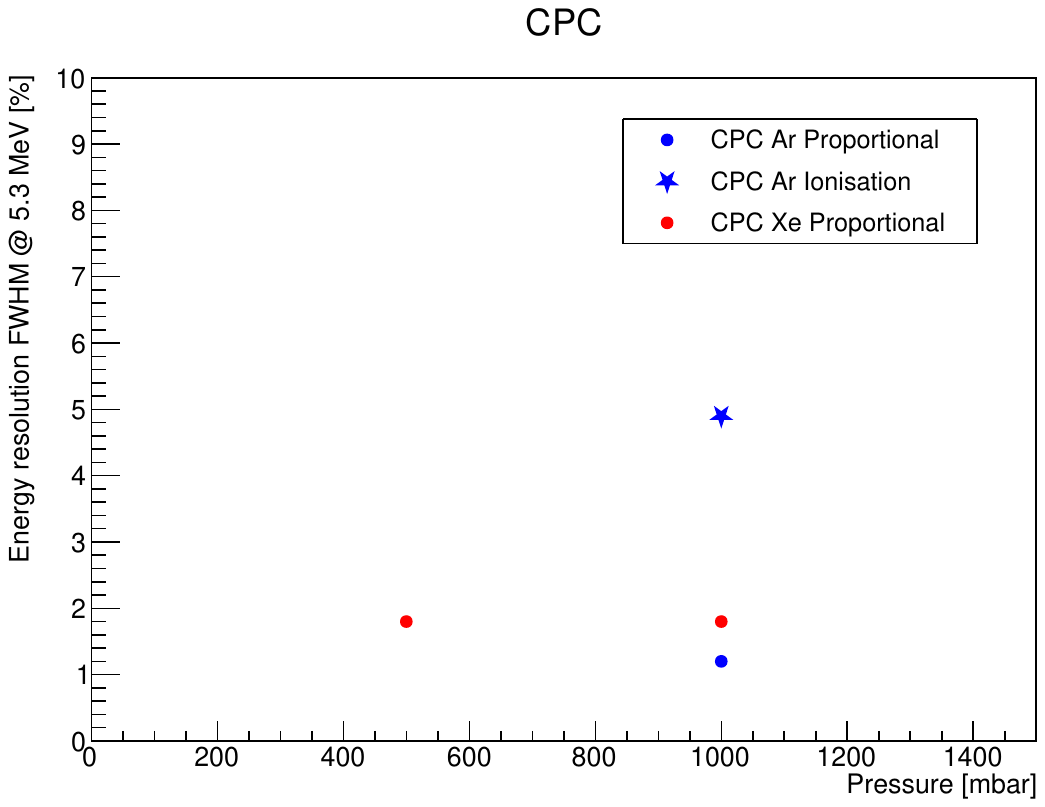}} \caption{{\it Energy resolution FWHM for 5.3 MeV alphas for SPC~\subref{fig:ResSPC}  and CPC~\subref{fig:ResCPC} in argon and xenon, in ionisation and proportional mode..}}
\label{fig:ResSPCCPC}
\end{figure}

As a summary, all the experimental results obtained in the different configurations of the R2D2 R\&D are shown  in Tab.~\ref{tab:1} and in Fig.~\ref{fig:ResSPCCPC}.

\begin{table}[t]
\begin{center}
\begin{tabular}{|c|c|c|c|c|c|c|c|}
\hline
Gas & Setup &Anode (radius)  & Pressure & HV & Noise (ADU) & Gain & Resolution\\
\hline
\multirow{8}{*}{ArP2}  & \multirow{6}{*}{SPC} &  & 200 mbar & 800 &  4.3 &45 & 1.1\%\\
 & &  & 500 mbar & 1300 &  4.1 &34 & 1.1\%\\
 &  & 1~mm &  1000 mbar & 1900 &  4.2 &30 & 0.9\%\\
 & &  & 2000 mbar & 2700 &  4.5 &10 & 1.3\%\\
 & &  & 3000 mbar & 3900 &  4.8 &10 & 1.3\%\\\cline{3-8}
\ & &  3~mm & 1000 mbar & 700 &  4.4 &1 & 8.2\%\\\cline{2-8}
 & \multirow{2}{*}{CPC} &  \multirow{2}{*}{10~$\mu$m}  & 1000 mbar & 200 &  3.6 &1 & 4.9\%\\
& & & 1000 mbar & 900 &  3.9 &9 & 1.2\%\\

\hline
\multirow{4}{*}{Xe} & \multirow{2}{*}{SPC}  & \multirow{2}{*}{3~mm} & 250 mbar & 1300 &  4.5 &1.3 & 3.8\%\\
 &  &  & 900 mbar & 1300 &  4.4 &1 & 7.2\%\\\cline{2-8}
 
 &  \multirow{2}{*}{CPC} &  \multirow{2}{*}{10~$\mu$m}  & 500 mbar & 900 &  3.8 &20 & 1.8\%\\
 &  &  & 1000 mbar & 1200 &  3.9 &14 & 1.8\%\\
\hline

\end{tabular}
\caption{{\it Summary of different experimental setup configurations and corresponding resolution obtained. Note that the cosmic reduction analysis is not accounted for in the table. The baseline noise in terms of ADU as well as the approximative operation gain are also stated.}}
\label{tab:1}
\end{center}
\end{table}%

\section{Data/Simulation comparison}
\begin{figure} [t]
\centering
\subfigure[\label{fig:det121}]{\includegraphics[height=5.5
cm]{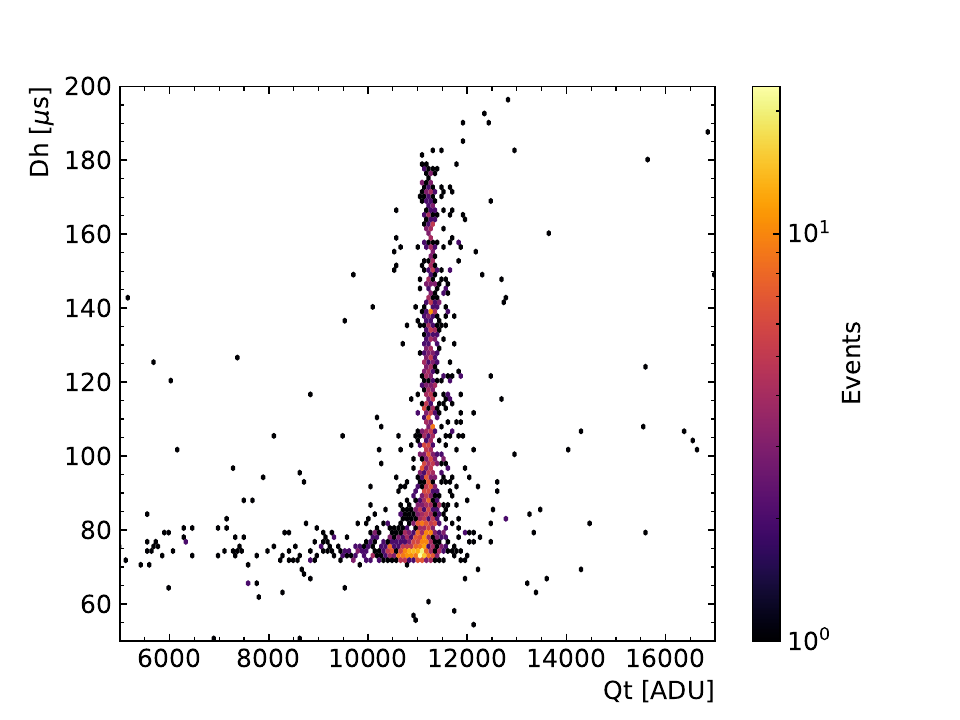}}
\subfigure[\label{fig:det122}]{\includegraphics[height=5.5
cm]{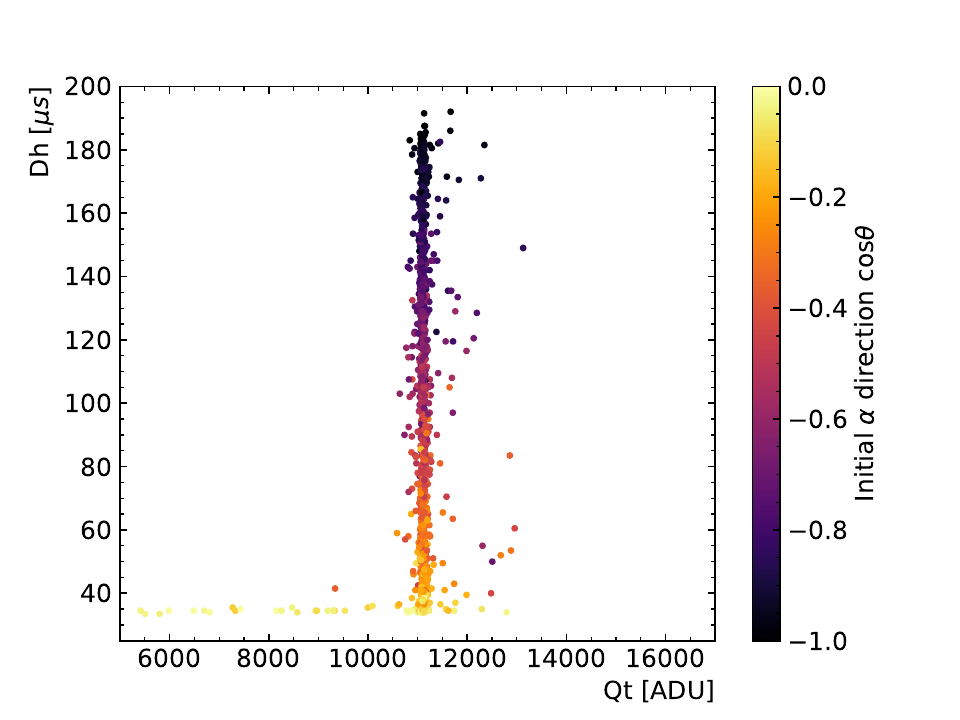}} \caption{{\it Signal width at half maximum ($Dh$) Vs. total charge ($Qt$) for data \subref{fig:det121}  and MC \subref{fig:det122} in xenon at 1~bar and 1200~V on the cathode. The color represents the number of events for data whereas for MC it indicates the direction of the $\alpha$ particle with respect to the wire.}}
\label{fig:12}
\end{figure}

Finally, the experimental observations were compared with the results of a Monte-Carlo simulation generating $\alpha$ particles of 5.3~MeV at a radius corresponding to that of the cathode. Such a comparison was carried out assuming in the MC a xenon pressure of 1~bar and an HV on the cathode of 1200~V to match the data taking conditions. The topology of the electric field, the energy deposition, and the diffusion of electrons during the drift were taken into account. The gas was assumed to be pure and no electronegative impurity was included in the simulation. The waveforms were reconstructed and compared to the collected data, in particular evaluating the signal width at half maximum ($Dh$) versus the collected charge ($Qt$).\\
The simulation allowed us to associate specific topologies of waveforms with the direction of $\alpha$ particles with respect to the central wire.  Particles going towards the wire exhibit larger signals which demonstrate that the signal width is driven mostly by the different drift time of electrons created at different radial positions, rather than by their diffusion. Conversely, particles tangent to the cathode, which produce electrons having the same drift time, generate a signal having a width dominated by the diffusion. The agreement between data and MC is very good for signals of $\alpha$ particles going towards the anode, whereas for particles tangent to the cathode the MC exhibits narrower signals as shown in Fig.~\ref{fig:12}. A better agreement can be obtained by artificially increasing the diffusion in the MC. Such a behaviour can be explained by the fact that impurities in the gas, not included in the simulation, could indeed result in the increase of the electron diffusion while drifting.

\section{Conclusions}
The R2D2 project aims to exploit radial TPCs, filled with gaseous xenon at high pressure, to build a tonne-scale detector for the search for $\beta\beta0\nu$. A key element of such  a technology is the possibility to  reach a resolution at the level of 1\% FWHM at 2.458~MeV, the $Q_{\beta\beta}$ of $^{136}$Xe.\\
The new setup operated at LP2I Bordeaux confirmed the previous results obtained in argon up to a pressure of 3~bar. For the first time the detector was operated in xenon, yielding excellent results up to 1~bar at the level of 1.4\% for $\alpha$ particles at 5.3~MeV. Two different detector geometries were tested and compared and the detector fonctionning was validated with the current setup. Determining the possibility of reaching the desired gas purity at higher pressure, without degrading energy resolution and the two track reconstruction, with a cylindrical TPC, is the final step of R\&D before moving toward a real-scale experiment.

\section*{Acknowledgments}
The authors would like to thank the IdEx Bordeaux 2019 Emergence program for the OWEN grant for the ``Development of a custom made electronics for a single channel time projection chamber detector aiming at the discovery of neutrinoless double beta decays, and for possible applications in industry''. We thank the LP2I Bordeaux and SUBATECH technical staff. 


\bibliographystyle{ieeetr}
\bibliography{references}

\end{document}